\documentclass[12pt]{article}
\usepackage{graphicx}
\usepackage{amssymb}
\usepackage{xcolor}

\def\eps{\epsilon}

\def\la{\lambda}
\def\si{\sigma}

\def\la{\lambda}
\def\vp{\varphi}

\def\ra{\Rightarrow}

\def\mm{\mathfrak{m}}

\def\cO{{\mathcal O}}

\def\11{{\mathbb 1}}

\def\rd{{\rm d}}

\def\MPL{M_{\rm Pl}}

\def\tPL{t_{\rm Pl}}

\def\MBPS{M_{\rm BPS}}

\def\beq{\begin{equation}}
\def\eeq{\end{equation}}
\def\bea{\begin{eqnarray}}
\def\eea{\end{eqnarray}}
\def\nn{\nonumber}

\def\ra{\rightarrow}

\begin{document}
\begin{center}
{\bf\Large Origin and growth of primordial black holes}\\[5mm]
{\bf Krzysztof A. Meissner$^1$ and Hermann Nicolai$^2$}\\[5mm]

{$^1$Faculty of Physics,
University of Warsaw\\
Pasteura 5, 02-093 Warsaw, Poland\\
$^2$Max-Planck-Institut f\"ur Gravitationsphysik\\
(Albert-Einstein-Institut)\\
M\"uhlenberg 1, D-14476 Potsdam, Germany\\
}
\vspace{3mm}
\end{center}

\begin{abstract} 
\noindent
Building on the insight that primordial black holes can arise from the 
formation and subsequent gravitational collapse of bound states of stable supermassive 
elementary particles during the early radiation era, 
we offer a comprehensive picture describing the evolution and growth of 
the resulting mini-black holes through both the radiation and matter dominated 
phases, until the onset of (small scale) inhomogeneities. This is achieved by means of an 
exact metric solving Einstein's equations throughout both phases.
We show that, thanks to a special enhancement effect producing an effective horizon 
above the actual event horizon, this process can explain the observed mass values of 
the earliest giant black holes. Unlike other proposals, it also predicts a lower 
limit on the mass of supermassive black holes. 
\end{abstract}

\vspace{5mm}
\noindent
\section{Introduction}

In very recent work a new mechanism was proposed to explain the origin 
of supermassive black holes in the early Universe by means of the condensation 
of superheavy elementary particles during the early radiation phase \cite{MN}.
Accordingly, the existence of primordial black holes would be due to  
the gravitational collapse of such bound states, shortly after their formation,
to small black holes, whose masses must lie above a certain critical value 
to evade Hawking evaporation.  Their subsequent growth during the radiation 
era can be modeled by an exact metric solving Einstein's equation, such 
that towards the end of the radiation era the emerging macroscopic black 
holes can grow to nearly solar mass objects.

In this Letter we discuss the complete evolution of such primordial black holes 
throughout both the radiation and matter dominated eras, and show 
that the proposed mechanism can indeed explain the observed mass 
values of supermassive black holes, as reported in \cite{GiantBH}. This completes
the argument given in \cite{MN}, where we did not follow the evolution of 
the emergent macroscopic black holes beyond equilibrium time $t_{eq}$,
and did not provide mass estimates for the large black holes that emerge at the time 
of the formation of small scale inhomogeneities. Here we close this crucial gap 
by offering a much more comprehensive picture, modeling the growth
of mini-black holes into giant black holes `from beginning to end'. The fact that 
this can be done by means of a closed form metric solving the Einstein equations 
that encompasses {\em both} the radiation and the matter dominated phase is an 
important input in our analysis.

As we have explained in \cite{MN}, superheavy gravitinos can serve as microscopic
seeds for generating mini-black holes if their mass is sufficiently large 
so that their gravitational attraction exceeds the repulsive or attractive 
electric forces between them. Furthermore, these seed particles must
be stable against decay into Standard Model matter. Although other kinds of 
particles with similar properties might serve the same purpose, we have argued in
\cite{MN} that the gravitinos of maximal ($N=8$) supergravity are distinguished 
in view of a possible unification of the fundamental interactions (however, as 
explained there, the underlying theory must transcend $N=8$ supergravity). 
This follows from the structure of the fermionic sector of the maximal $N=8$ 
supermultiplet \cite{MN0}: identifying  the 48 non-Goldstino spin-$\frac12$ fermions 
of the $N=8$ supermultiplet with three generations of quarks and leptons 
of the Standard Model of particle physics (including right-chiral neutrinos),
one is left with eight massive gravitinos
with the properties described in \cite{MN0,MN}. These properties are radically different from
those of the more familiar sterile gravitinos of low energy $N=1$ supergravity models; in particular,
unlike the latter, superheavy gravitinos {\em do} participate in Standard Model interactions. 

Although our proposal is thus mainly motivated by unification, we emphasize again that,
except for the properties listed below in section 2, our considerations are largely 
independent of the precise nature of the ``seed particles" that produce primordial black holes.
Evidently, our proposal differs in several important ways from other scenarios aiming to 
explain the origin of primordial black holes, which we cannot review here for lack of space.
See, however \cite{Kh,V} for alternative ans\"atze, and \cite{CKSY} for a comprehensive survey 
of the present state of the art and a discussion of the relative merits of different proposals.

\section{Basic considerations}

We refer readers to \cite{MN} for a more detailed explanation of the basic 
motivation and assumptions underlying our proposal.
As argued there, the gravitino mass $M_g$ must lie
between $\MBPS$ and $\MPL$, where the `BPS-mass' $\MBPS$ is the mass for 
which the electrostatic repulsion between two (anti-)gravitinos of the same charge equals 
their gravitational attraction.
$\MPL$ is the reduced Planck mass $\sim 4.34\cdot 10^{-9}$ kg (it corresponds
to the Planck time  $\tPL = 2.70\cdot 10^{-43}\,$s). For numerical estimates we 
will take $\MBPS \sim 0.01 \cdot \MPL$, so that
\beq\label{Mg}
0.01 \cdot \MPL \, < \,  M_g \,<\, \MPL
\eeq
This ensures that the force remains attractive also between gravitinos of the 
same electric charge. 
The minimal seed mass $M_{seed}\sim N M_g$ for a primordial black hole
in the early radiation phase is determined by  asking the total 
energy of a bound system of $N$ (anti-)gravitinos to be negative, {\it viz.}
\footnote{We adopt units with $\hbar = 1$, $c=1$, $k_B = 1$, so
 that for instance 1 eV = 1.16$\,\cdot\,10^4\,$K, {\em etc.} The final formulas
are then re-expressed in convenient units (eV, kg, m, s, or K).}
\beq
\langle E_{kin}(t) \rangle + \langle E_{pot}(t) \rangle \,=\,
N T_{rad}(t) - N^2 \frac{GM_g^2}{\langle d(t) \rangle} \, \stackrel{!}{<} \, 0
\eeq
where $\langle d(t)\rangle$ is the (time-dependent) average distance between 
two gravitinos in the ambient hot radiation plasma.
As we explain in  \cite{MN}, the cosmic time $t$ drops out in this inequality
upon substituting the relevant quantities with their time dependence.
We then find
\beq\label{N}
N \,\gtrsim \,  \frac{T_{eq}}{GM_g^2} \cdot 10^2\,{\rm m} \,\sim\, 10^{12} 
\eeq 
for the minimum number of (anti-)gravitinos in a bound state for gravitational
collapse to occur, where $T_{eq} \sim 1\,$eV and we take $M_g \sim 10^{-9}\,$kg 
as an exemplary value \cite{MN}. Since the cosmic time $t$ drops out in the derivation of this 
inequality, the value of $N$ remains the same throughout the radiation phase. 
If the bound state is  meta-stable, the collapse can be delayed 
in such a way that an even larger number  $N$ of  (anti-)gravitinos can 
accrue before gravitational collapse occurs, in which case the seed mass 
could be even larger. The minimum mass of a black hole resulting from gravitational 
collapse of such a bound state is therefore
\beq\label{Mlump}
M_{seed} \,\sim\, 10^{12} M_g   \, \sim \, 10^3\, {\rm kg} \;
\Rightarrow \, GM_{seed} \sim 10^{-24}\,{\rm m}
\eeq
Now, a black hole of such a small mass would be expected to decay immediately 
by Hawking radiation \cite{Wald}: from the well known formula for the lifetime of a 
black hole (see {\em e.g.} \cite{TD}) we have
\beq\label{BHtemp}
\tau_{evap} (m) \,=\, \tPL \left( \frac{m}{\MPL} \right)^3
\eeq
This is the result which would hold in empty space. However, during the early
radiation phase this is not the only process that must be taken into account, because
of the presence of extremely hot and dense radiation, which 
can `feed'  black hole growth. The absorption of radiation thus provides 
a competing process which can stabilize the black hole against Hawking decay, such
that with the initially extremely high temperatures of the radiation era mass accretion 
can overwhelm Hawking evaporation {\em even for very small black holes}. 
The details of this process are complicated, because a proper treatment would
require generalizing the original Hawking calculation
to the time-dependent space-time background given by (\ref{MDmetric}) below,
something that remains to be done. However, there is a simple approximate 
criterion for accretion to overcome the rate for Hawking radiation 
for a black hole of given mass $m$, which reads
\beq\label{Trad}
T_{rad} (t) \,>\,  T_{Hawking}(m)  \,=\, \frac{1}{8\pi  Gm}  
\eeq
The break-even point is reached when the radiation temperature equals the
Hawking temperature, at time $t_0 = t_0(m)$ when $T_{rad}(t_0) \sim T_{Hawking}(m)$.
For larger times  $t > t_0$ (and lower radiation temperatures) a black hole of mass $m$ will decay.
Imposing this equality, or alternatively using eqn.(26) of \cite{MN} we deduce 
the relevant mass at time $t$, which gives
\beq\label{even}
m^4(t) \,\simeq \, \frac{\MPL^3}{\tPL} \cdot \frac1{G^2 \rho_{rad}(t)} \,=\,
 \frac{32\pi \MPL^3}{3G \tPL} \cdot t^2
\eeq
When read from right to left this equation tells us which is the latest time for a mini-black 
hole of given mass $m$ to remain stable against Hawking decay during the radiation phase. 
This is the case  for $t< t_0 \equiv t(m) \propto m^2$, after which time the 
black hole will decay. Conversely, for a given time $t$ 
any mini-black hole of initial mass greater than $m(t)$ will be able to survive 
and can start growing, whereas those of smaller mass decay. With (\ref{Mlump}) 
as the reference value we thus take the initial mass to be $\sim M_{seed}$, and 
assume that the time range available for the formation of such a mini-black hole is
\beq\label{tseed}
t_{min}=10^8 \cdot \tPL \simeq 10^{-34}\,{\rm s}\;<\; t \;<\;  t_{max} \simeq 10^{-18}\,{\rm s}
\eeq
During this time interval a black hole of initial mass (\ref{Mlump}) can survive 
and start growing by accreting radiation. While the upper bound is thus
determined by setting $t_{max} \equiv t(M_{seed})$, the lower bound has been 
chosen mainly  to stay clear of the quantum gravity regime and a possible
inflationary phase.

Once we have a stable mini-black hole  we can study its further evolution through
the radiation phase by means the exact solution derived in \cite{MN}, until matter starts
to dominate over radiation at time  $t \sim t_{eq} \sim 42000\,$yr, when these objects
have grown into macroscopic black holes.
With (\ref{tseed}) we get the following range of masses 
\beq
10^{-12}\ M_\odot\,\lesssim \, m(t_{eq}) \,\lesssim \, 10^{-3}\ M_\odot
\eeq
However, the solution in \cite{MN} does {\em not} apply to the matter dominated phase. 
To investigate the further evolution one would conventionally switch to a different description 
by invoking the Eddington formula \cite{GiantBH,Eddington}
\beq\label{Edd1}
m(t)=M_0\,\exp\left(\frac{4\pi G m_p t}{\eps \si_T}\right)\simeq
 M_0\,\exp\left(\frac{ t}{45\, {\rm Myr}}\right)
\eeq
where $m_p$ is the proton mass, $\si_T$ is the Thompson cross section, and $\eps$ 
is the fraction of the mass loss that is radiated away (usually taken as  $\eps=0.1$).  
Unfortunately, because of the exponential dependence this formula
is extremely sensitive to the precise value of $\eps$ and the choice of ``final"
time $t$ -- surely, exponential growth does not persist into the present epoch!

We also note that this formula was originally developed to describe the 
evolution of luminous stars \cite{Eddington}. Its derivation relies 
on the Newtonian approximation and is based on a simple equilibrium condition, 
balancing the rate of mass absorption against the luminosity of infalling matter, 
where the luminosity is assumed to grow linearly with the mass of the star. 
It thus appears doubtful whether one can use it in the present context, and 
we therefore prefer to refrain from a `blind' application of (\ref{Edd1}).
Instead we here propose a general relativistic treatment of black hole evolution in a dense 
environment by means of an exact solution of Einstein's equations, which seems 
superior to (\ref{Edd1}) even though it does not (yet) take into account rotation and matter self-interactions.
Furthermore, unlike (\ref{Edd1}), our final formula does not rely on exponential growth.

\section{Black hole evolution from radiation to matter dominated era}

To present this new solution we employ conformal coordinates, with conformal 
time $\eta$, instead of the cosmic time coordinate $t$ used above. 
One main advantage of this coordinate choice is that the causal structure 
of the space-time is often easier to analyze (for the solution to be presented below it is the 
same as that of the Schwarzschild solution). Secondly, we wish to exploit the 
fact that the use of conformal time allows us to exhibit a simple closed form solution 
that encompasses {\em both} the radiative and the matter dominated phase.
With conformal time $\eta$, the Friedmann equations 
read (for a spatially flat universe and vanishing cosmological constant)
\beq\label{Friedmann}
\dot a^2 \,=\,  \frac{8\pi G}{3} \rho a^4 \;\; ,\;\;
a \ddot a - \dot a^2 \,=\, - \frac{4\pi G}3 (\rho + 3p) a^4
\eeq
where 
\beq
\dot a \equiv \frac{\rd a}{\rd\eta} \;\;,\quad
\rd t = a(\eta) \rd\eta   \;.
\eeq 
The requisite exact solution of (\ref{Friedmann}) is (see {\em e.g.} \cite{MFB}).
\beq\label{aRDMD}
a(\eta) = A\eta + B^2 \eta^2  \;\;  \qquad \Big( \Rightarrow \
t = \frac12 A\eta^2 + \frac13 B^2 \eta^3 \, \Big)
\eeq
together with the density and pressure  
\bea\label{rhop1}
8\pi G \rho(\eta) &=&  \frac{3A^2}{a^4(\eta)} + \frac{12 B^2}{a^3(\eta)} \quad ,\quad
8 \pi G p(\eta) = \frac{A^2}{a^4(\eta)}   
\eea
The relevant numbers $A$ and $B$ can be calculated from known data, up to rescaling
$\eta\ra\la\eta,\ A\ra\la^{-2}A,\  B\ra \la^{-3/2}B,\  a\ra \la^{-1} a$. The latter scale is
conventionally fixed by setting $a(t_0)\!=\!1$, where $t_0\simeq 13.8\cdot 10^9\,$yr
is the present time. Taking this as the reference value we make use
of the fact that at equilibrium between radiation and matter \cite{Pl}
\beq\label{NSeq}
a(\eta_{eq})\simeq\frac{1}{3400},\ \ \ \  
t_{eq}\simeq 1.5\cdot 10^{12}\ {\rm s}
\eeq
and at the last scattering \cite{Pl}
\beq\label{LSeq}
a(\eta_{LS})\simeq\frac{1}{1090},\ \ \ \  
t_{LS}\simeq 1.2\cdot 10^{13}\ {\rm s}
\eeq
This gives
\beq\label{AB}
A= 2.1\cdot 10^{-20}\ {\rm s}^{-1},\ \ \ \ B = 6.2\cdot 10^{-19}\ {\rm s}^{-1}.
\eeq
for our Universe (starting from nucleosynthesis).

For the new metric ansatz we now substitute (\ref{aRDMD}) into~\footnote{This
 ansatz is somewhat similar to, but actually different from, the McVittie solution
\cite{McV0,McV1,McV2,McV3,McV4} and the Lema\^{\i}tre-Tolman-Bondi metric \cite{Kr}.
This follows for instance from the fact that for our solution the black hole mass
grows with time, cf. (\ref{mm}) below, and has a non-vanishing heat flow vector 
$q_\mu \neq 0$ in (\ref{Weinberg}).}
\beq\label{MDmetric}
\rd s^2\,=\, a(\eta)^2\left[- \tilde{C}(r)\rd \eta^2+ \frac{\rd r^2}{\tilde{C}(r)}+r^2\rd\Omega^2\right]
\eeq
Here the {\em a priori} unknown function $\tilde{C}(r)$ is uniquely fixed by 
imposing two physical requirements corresponding to  the two limiting cases 
of pure matter and pure radiation. For pure radiation ($B=0$) we demand the 
trace of the energy-momentum tensor resulting from (\ref{MDmetric}) to vanish
\beq
T^\mu{}_\mu\,\stackrel{!}{=} \,0 \quad \Rightarrow  \qquad 
\frac{\rd^2}{\rd r^2} \big(r^2\tilde{C}(r) \big)  \,\stackrel{!}{=} \, 2 \; .
\eeq
With the standard form of the energy-momentum tensor for a perfect fluid
({\em i.e.} (\ref{Weinberg}) below for $q_\mu = 0$),
this is equivalent to the statement that $\rho = 3 p$ throughout the 
radiation era.  For the other limiting case of pure matter ($A=0$), we require the  
pressure to vanish: $p=0\Rightarrow (r\tilde{C})' \stackrel{!}{=}1$.  
This leads to the unique solution
\beq\label{C}
\tilde{C}(r) \,\equiv \,C(r): = 1 - \frac{2G\mm}{ r}
\eeq
which we will use in the following. The essential new feature here is that the 
metric (\ref{MDmetric}) allows us to evolve the black hole through {\em both} 
the radiative and matter dominated periods, with a smooth transition between the two.

In (\ref{C}) we use a different font for the fixed mass parameter
because $\mm$ is {\em not} the physical mass, unlike $m(t)$ above.
This is most easily seen by replacing
\beq\label{mm}
\frac{G\mm}{r} \;\; \rightarrow \;\; \frac{G\mm a(\eta)}{r a(\eta)} \,\equiv \,
        \frac{G\mm a(\eta)}{r_{phys}}    \quad\Rightarrow \quad 
        m(\eta) = \mm a(\eta)
\eeq
Using (\ref{Mlump}), (\ref{tseed}) and the above relation with $\eta_{min} = 10^{-7}\,$s
and $\eta_{max}= 10\,$s, as well as $G\mm_{min} = GM_{seed}/a_{max}$ 
and  $G\mm_{max} = GM_{seed}/a_{min}$ we get
\beq\label{mmminmax}
G \mm_{min}\sim 5\cdot 10^{-6}\ {\rm m},\ \ \ \ G \mm_{max}\sim 5\cdot 10^{2}\ {\rm m}
\eeq
For the metric ansatz (\ref{MDmetric}) with $C(r)$ from (\ref{C}) the non-vanishing 
components of the Einstein tensor,  hence the associated energy-momentum tensor, 
are given by:
\bea\label{Tmn2}
8\pi G \, T_{\eta\eta} &=& \frac{3\,\dot a^2}{a^2} \,=\, 
\frac{3(A+2B^2\eta)^2}{(A\eta + B^2 \eta^2)^2}
\nn\\[2mm]
8 \pi G \, T_{r\eta} &=& 
\frac{2G\mm}{r^2 C(r)} \cdot \frac{\dot a}{a}  \,=\, 
\frac{2G\mm}{r^2 C(r)}\! \cdot\! \frac{A+ 2B^2\eta}{A\eta + B^2\eta^2} \nn\\[2mm]
8\pi G T_{rr} &=& \frac{\dot a^2 - 2a\ddot a}{a^2 C(r)^2 }
\, = \, \frac{1}{C(r)^2}\! \cdot\! \frac{A^2}{(A\eta + B^2 \eta^2)^2}
\eea
together with~\footnote{We take this 
  opportunity to correct two misprints in \cite{MN}: the extra factor of $C$ in  
  (\ref{T}) below is missing in (46) there. Furthermore, in eqn.(50) of \cite{MN} it should read
  $$
 8\pi G p(\eta,r) \,=\, \frac{r}{A^2 \eta^4 (r- 2Gm)} 
  $$
}
\beq\label{T}
T_{\theta\theta} \,=\,  C(r)\, r^2\, T_{rr} \;\; , \quad
T_{\vp\vp} \,=\, \sin^2\theta\,T_{\theta\theta}
\eeq	

Now, to elevate (\ref{Tmn2})  beyond the status of a mere identity, we must endow 
it with physical meaning by interpreting the r.h.s. in terms of physical sources of energy 
and momentum, that is, a proper energy-momentum tensor, appropriate to radiation
and matter.  To this aim we re-express 
the r.h.s. of (\ref{Tmn2}) in the standard form \cite{Weinberg}
\beq\label{Weinberg}
T_{\mu\nu} = p g_{\mu\nu} + (p +\rho)\, u_\mu u_\nu - u_\mu q_\nu - u_\nu q_\mu
\eeq
Here we neglect higher derivatives in $u_\mu$ and matter self-interactions
(viscosity, {\em etc.}). For the density and pressure to match between (\ref{Weinberg})
and (\ref{Tmn2}) we must include an extra inverse factor $C(r)$  in comparison 
with (\ref{rhop1}) to account for the curvature
\bea\label{rhop2}
8\pi G \rho(\eta,r) \,&=&\,  \frac1{C(r)} \left(\frac{3A^2}{a^4(\eta)} + \frac{12 B^2}{a^3(\eta)}\right)
 \nn\\[2mm]
8 \pi G p(\eta,r) &=& \frac1{C(r)}\frac{A^2}{a^4(\eta)}   
\eea
again with $a(\eta)$ from (\ref{aRDMD}). The 4-velocity is  
\footnote{There is a second solution with the same
 $\rho$ and $p$, but $u_r = q_\eta = 0$, 
 which we discard as unphysical because it would imply the absence of
 motion of matter other than the co-motion 
  with the cosmic frame.}
\beq\label{u}
u_\mu\,=\, - \frac{a(\eta)}{ C(r)^{1/2} }\,\big(C(r)\cosh\xi\,,\, \sinh\xi \,,\,0 \,,\,0\big)
\eeq
while the heat flow vector is given by
\beq\label{qq}
8\pi G
q_\mu \,=\, -\frac{2 G\mm {\dot a(\eta)}}{r^2 C(r)^{3/2} a(\eta)^2}\, \big(C(r)\sinh\xi \,,\,\cosh\xi \,,\,0 \,,\,0)
\eeq
These vectors obey $u^\mu u_\mu = -1$ and $u^\mu q_\mu = 0$. 
The parameter $\xi  = \xi(\eta,r) > 0$ is determined from
\beq\label{xi}
\tanh\xi \,=\, \frac{G\mm\eta}{r^2}\cdot\left(1-
\frac{ B^4\eta^2}{A^2 + 3AB^2\eta + 3B^4\eta^2}\right)
\eeq
The signs in (\ref{u}) and (\ref{qq}) are chosen such that for the contravariant components of the 
4-velocity we have $u^\eta > 0$ and $u^r < 0$, hence {\em inward} flow of matter. 
(Choosing the opposite sign for the components of $u_\mu$ would correspond to a 
shrinking white hole.)

To keep $\xi$ real and finite we must demand $\tanh\xi < 1$. It is readily seen that
\bea\label{tanh}
\tanh\xi &\sim&  \frac{G\mm \eta}{r^2}  \qquad \;
     \mbox{for $B^2\eta \ll A\quad$ (radiation)} \nn\\[1mm]
      &\sim&  \frac23 \frac{G\mm \eta}{r^2}  
     \;\;\quad \mbox{for $B^2\eta\gg A\quad$ (matter)} 
\eea
The representation (\ref{Weinberg}) is valid as long as all quantities remain 
real and finite. This requires $r^2 \,>\, \cO(1) G\mm \eta$, with a strictly positive 
$\cO(1)$ prefactor. When $r$ reaches the value for which $\tanh\xi =1$ the components of 
$u_\mu$ and $q_\mu$ diverge, and the expansion (\ref{Weinberg}) breaks down. 
For the external observer the average velocity of the infalling matter then reaches 
the speed of light, so for all practical purposes everything  happening inside this shell 
is shielded from the outside (even though light rays can still 
escape from this region, as long as $r > 2G\mm$). 
As we  are not concerned with $\cO(1)$ factors here we define
 \beq\label{rh}
r_H(\eta)  \, :=\,  a(\eta) \sqrt{{G\mm\eta}}
\eeq
and interpret the associated outward moving shell as an {\em effective} horizon 
(or `pseudo-horizon') that lies above the actual event horizon; note that $r_H(\eta)$ 
is invariant under the coordinate rescalings mentioned after (\ref{AB}).
Physically, we expect the matter inside the shell $r_{phys} \lesssim r_H(\eta)$ to be 
rapidly sucked up into the black hole, once the outside region $r_{phys} > r_H(\eta)$
gets depleted of `fuel' due to the formation of  inhomogeneities.  
The extra matter inside the shell $r_{phys} \lesssim r_H(\eta)$ thus enhances the growth
substantially, beyond the linear growth with the scale factor implied by (\ref{mm}). 

At the onset of inhomogeneities, we must stop using the metric (\ref{MDmetric}) because
the growth of the black hole gets decoupled from the growth of the scale factor $a(\eta)$,
after which the black hole evolves in a more standard fashion by much slower 
accretion (for this reason there is also no point in extending the metric 
ansatz (\ref{MDmetric}) into the present epoch, which is dominated by Dark Energy).
To estimate its mass we take the value of $r_H$ at that particular time to define 
an effective Schwarzschild radius, thus equating the mass with the maximum
energy that can possibly fit inside a shell of radius $r_H$. This approximation appears 
justified not only because of the apparent divergent kinetic energy of the infalling matter 
near $r_H$, but also because of the strong increase of the density and pressure 
inside this shell, due to the extra factor $C^{-1}(r)$ in (\ref{rhop2}). 
A more detailed investigation of the evolution inside the shell in view of 
eliminating the firewall at $r=2G\mm$ would require modifying the metric 
ansatz (\ref{MDmetric}) for $r \lesssim r_H(\eta)$, for instance replacing $C(r)$ by $C(r,\eta)$.

\section{Mass Estimates}

We can now apply the above formulas to estimate the resulting black hole mass at
the onset of (small scale) inhomogeneities, {\em i.e.} the onset of star formation. 
To be sure, there are still uncertainties about the actual numbers, but it is reassuring
that we do end up the right orders of magnitude. 
The relevant time $t$ at which to evaluate $ r_H(\eta(t))$ lies well after decoupling, 
since the inhomogeneities in the CMB are still tiny, of order $\cO(10^{-5})$. 
Rather, we take $t_{inhom}\simeq 10^8\,$yr $\simeq 3.2\cdot 10^{15}\,$s, 
which is the time when the first stars are born \cite{OldStar}. This corresponds to  
$ \eta_{inhom} \simeq 2.7\cdot 10^{17} \,{\rm s} \Rightarrow 
  a(\eta_{inhom}) \simeq 0.034$.
Substituting (\ref{mmminmax}) into (\ref{rh}) and using $r_S(M_\odot)=3$ km we can calculate 
the range of possible black hole masses, for instance taking $t\sim 100$ Myr as an 
approximate reference value: 
\beq\label{bounds}
10^{5}\ M_\odot \,\lesssim\,  m_{\rm BH}\,\lesssim\,  2\cdot 10^9\,M_\odot  
\eeq
which is consistent with observations \cite{GiantBH}. To reach such large mass values the replacement
of $G\mm$ by $\sqrt{G\mm\eta}$ in (\ref{rh}), as advocated in this paper, is evidently of crucial importance.

Observe that, as a consequence of the Hawking evaporation of seed black holes with
too small mass, our calculation also provides a {\em lower bound} in (\ref{bounds}),
in contradistinction to other proposals where there is no such
lower bound on the mass range. It is thus a prediction of the present mechanism that the black holes formed from gravitinos should belong to a very different mass category than 
the black holes formed from stellar collapse and subsequent mergers, a prediction that can
also serve to discriminate our proposal against alternative ones. 
From the present point of view, the existence of such a gap in the mass distribution 
of black holes in the Universe would thus constitute indirect observational 
evidence for the existence of Hawking radiation. At least so far, this expectation 
seems to be in accord with observation, as no such objects (intermediate mass black holes
= ``IMBHs") have been found until now \cite{VM}.

\newpage

\noindent
{\bf Note added:} After this paper was accepted for publication we became 
aware of early work \cite{Th} which discusses a metric ansatz similar to (\ref{MDmetric}).
Possible astrophysical applications different from the ones considered 
here have very recently been considered  in \cite{P}.

\vspace{0.3cm} 
\noindent
 {\bf Acknowledgments:} 
 We thank the referee for comments that helped to improve this paper.
K.A.M. was partially supported by the Polish National Science Center grant.
The work of  H.N. has received funding from the European Research 
Council (ERC) under the  European Union's Horizon 2020 research and 
innovation programme (grant agreement No 740209).  The title of this paper was
inspired by a philosophical treatise by one of the first author's ancestors \cite{WL}.

\end{document}